\pdfoutput=1

\documentclass[11pt]{article}

\usepackage[final]{acl}
\usepackage{times}
\usepackage{latexsym}

\usepackage[T1]{fontenc}

\usepackage[utf8]{inputenc}
\usepackage{amssymb}
\usepackage{pifont}
\usepackage{microtype}
\usepackage{inconsolata}
\usepackage[most]{tcolorbox}
\usepackage{graphicx}
\usepackage{array}
\usepackage{hyperref}
\usepackage[table]{xcolor}  
\usepackage{colortbl}        

\usepackage{tikz}
\usepackage{tabularx}
\usepackage{subfigure,graphicx}
\usepackage{booktabs}
\usepackage{multirow}
\usepackage{makecell}
\usepackage{xcolor}
\usepackage{eqparbox} 
\usepackage{transparent}

\newcommand{\eg}{\textit{e.g.}}

\newcommand{\ours}{Laser}

\title{
\ours{}: Governing Long-Horizon Agentic Search via Structured Protocol and Context Register}

\author{Shuting Wang$^{1,2}$, Qiaolin Xia$^{1}$, Vich Wang$^{1}$, Herberttli$^{1}$, Bobsimons$^{1}$, \textbf{Zhicheng Dou$^{2*}$} \\
$^1$ Tencent Hunyuan  \\
$^2$ Gaoling School of Artificial Intelligence, Renmin University of China  \\
\texttt{\{wangshuting\}@ruc.edu.cn}
}
\begin{document}

\maketitle
\def\thefootnote{*}\footnotetext{Corresponding author.}
\begin{abstract}
Recent advances in Large Language Models (LLMs) and Large Reasoning Models (LRMs) have enabled agentic search systems that interleave multi-step reasoning with external tool use. However, existing frameworks largely rely on unstructured natural-language reasoning and accumulate raw intermediate traces in the context, which often leads to unstable reasoning trajectories, context overflow, and degraded performance on complex multi-hop queries.
In this study, we introduce \ours{}, a general framework for stabilizing and scaling agentic search. \ours{} defines a symbolic action protocol that organizes agent behaviors into three spaces: planning, 
task-solving, 
and retrospection. 
Each action is specified with explicit semantics and a deterministic execution format, enabling structured and logical reasoning processes and reliable action parsing. This design makes intermediate decisions interpretable and traceable, enhancing explicit retrospection and fine-grained control over reasoning trajectories.
In coordination with parsable actions, \ours{} further maintains a compact context register that stores only essential states of the reasoning process, allowing the agent to reason over long horizons without uncontrolled context expansion. 
Experiments on Qwen2.5/3-series models across challenging multi-hop QA datasets show that \ours{} consistently outperforms existing agentic search baselines under both prompting-only and fine-tuning settings, demonstrating that \ours{} provides a principled and effective foundation for robust, scalable agentic search.
\end{abstract}

\section{Introduction}\label{sec:intro}
\begin{figure}
    \centering
    \includegraphics[width=1\linewidth]{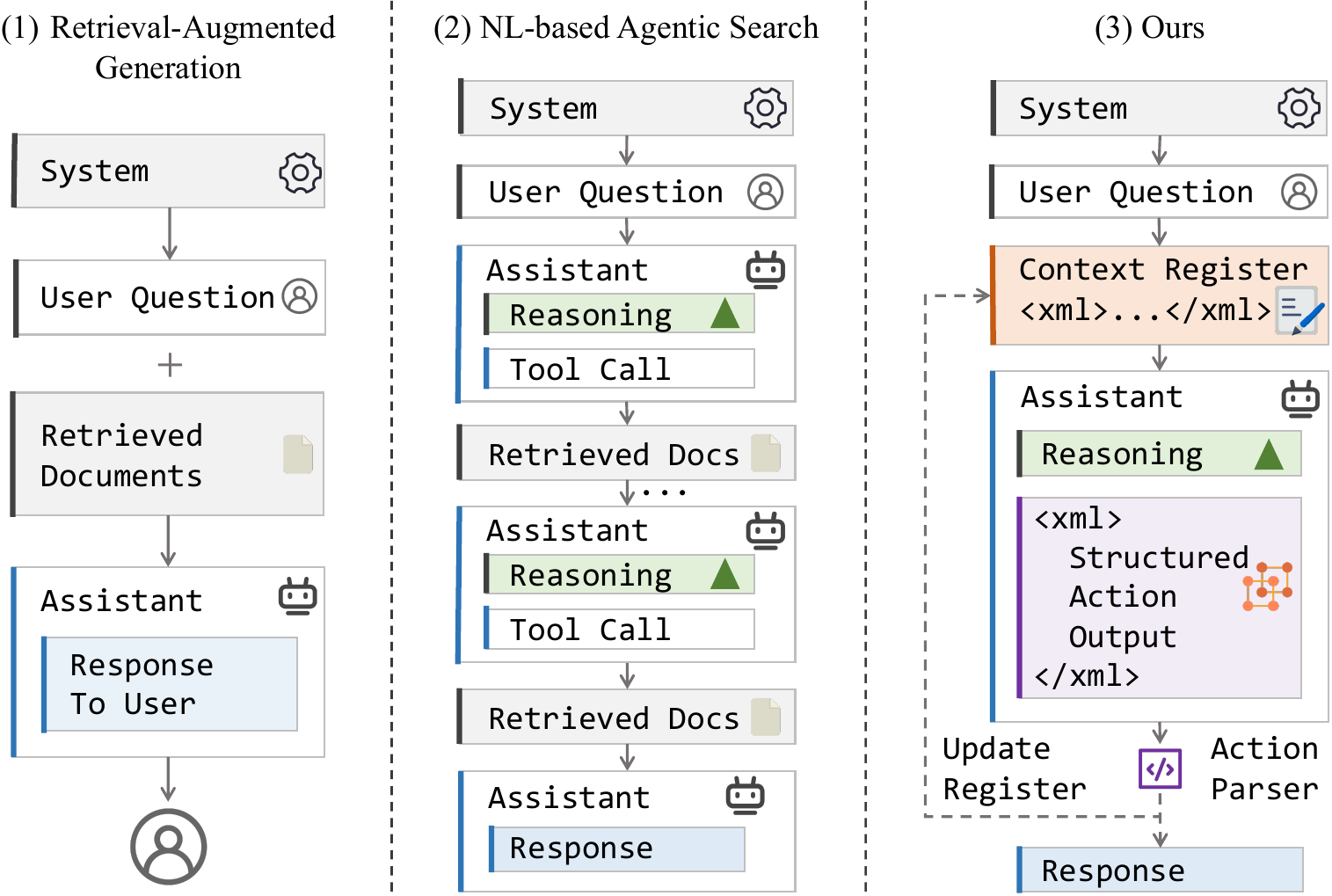}
    \caption{Visualization of the comparison between \ours{} and existing LLM-based AI search models.}
    \label{fig:intro}
\end{figure}

The advent of large language models (LLMs) has fundamentally transformed information access, moving from returning web pages to generating direct answers~\cite{rag,radit,RichRAG}. Recently, Large Reasoning Models (LRMs) with thinking and tool-calling abilities have been powering a new generation of agentic search systems~\cite{searchr1,research,r1searcher}. Unlike traditional search engines, search agents are expected to handle deeper multi-hop queries, gather information via tool-calling, and reason through ambiguous and complex user tasks. Frameworks such as ReAct~\cite{yao2023react} and Search-R1~\cite{searchr1} have empowered agents to interleave reasoning with action, allowing them to interact with external tools to solve problems dynamically.

However, as user queries become increasingly intricate, existing agentic frameworks face two critical bottlenecks: 
\textit{the limitations of natural language reasoning} and \textit{context overflow}.
First, most search agents rely on linear chains of free-form natural language (NL). While flexible, this modality lacks the symbolic rigor required to model complex, non-linear task structures. Without a rigorous structural guide, 
the reasoning process is prone to degenerating into unstructured loops or premature conclusions. Second, the accumulation of raw interaction logs (\eg retrieved knowledge and intermediate reasoning) rapidly consumes the context window, leading to ``context pollution'' where critical information is obscured by noise, ultimately degrading both reasoning quality and scalability.

To address these challenges, we introduce \ours{}, a novel framework designed to govern \textbf{L}ong-horizon \textbf{A}gentic search by introducing a \textbf{S}tructured protocol that regularizes the reasoning process and a cont\textbf{E}xt \textbf{R}egister that dynamically manages the evolving context. Together, these components ensure high-quality, efficient context usage while enabling structured, reliable reasoning over complex search tasks. 

Our design is inspired by classical theories of human problem solving, which emphasize structured problem representation, goal-directed execution, and metacognitive control~\cite{newell1972human,nelson1990metamemory}. 
Guided by these principles, \ours{} organizes the agent’s action space into three essential categories: planning, task-solving, and retrospection
, forming a functionally complete foundation for structured agentic reasoning. 
The planning space comprises actions of ``intent refinement'' and ``problem framing'', enabling fine-grained interpretation of user intent and graph-structured task decomposition;
The task-solving space includes ``tool call'', ``doc extraction'', ``task answer'', and ``final answer'', which collectively execute the core problem-solving procedure; 
The retrospection space introduces ``replanning'' and ``revisit-task'' actions, equipping the agent with reflective and corrective capabilities when intermediate solutions are found to be incomplete or incorrect.
Each action is defined with explicit semantics and a standardized executable format, ensuring that the reasoning process is governed by clear structural constraints rather than uncontrolled natural-language.

Moreover, this structured protocol enables reliable parsing of reasoning actions directly from the model’s responses through deterministic code, rather than through fragile generative models. Leveraging this complexity, \ours{} maintains a context register that stores only the essential state required for continued reasoning. This register serves as a condensed surrogate for the full reasoning history, allowing the agent to scale across many rounds of reasoning actions with a controlled, significantly reduced rate of context growth.
Figure~\ref{fig:intro} visualizes a comparison between our method and both RAG and NL-based agentic search systems.

We evaluate \ours{} on a range of open-source LLMs, including Qwen2.5-7B~\cite{qwen2.5} and Qwen3-8B/32B~\cite{qwen3}, as well as on challenging multi-hop question answering (QA) benchmarks such as BrowseComp-ZH~\cite{browsecomp-zh} and WebDancer~\cite{webdancer}.
Across all settings, \ours{} consistently outperforms current agentic search baselines, demonstrating the effectiveness of our structured protocol and context register. Additional analyses further highlight the contribution of each module and the significant reduction in context-token usage achieved by \ours{}. We open-source the code of our method at \hyperlink{https://github.com/ShootingWong/Laser.git}{https://github.com/ShootingWong/Laser.git}.

\section{Related Work}
\label{sec:related_work}
\subsection{Retrieval-Augmented Generation}
Retrieval-augmented generation (RAG) enhances LLM performance by grounding generation in external knowledge. Early work focuses on jointly or collaboratively optimizing retrievers and generators~\cite{atlas,retro,retroplus,realm,hindsight,rag,radit,self-rag}. 
To reduce training and deployment costs, subsequent studies fix the LLM backbone and improve retrieval through modular plug-ins~\cite{replug,ARR} or lightweight post-retrieval components, such as compressors and re-rankers~\cite{comp1,zhengbao_filter,xiong_filter,bgm,ragsurvey}. 
In parallel, multi-hop RAG methods decompose complex queries into intermediate sub-questions and retrieval steps~\cite{chan2024rqrag,wang2024selfdc,khot2023decomposed,searchchain,RichRAG}. 
However, these approaches are largely constrained by the foundational capacity of earlier-generation LLMs and typically rely on shallow or linear decomposition, limiting their effectiveness on long-horizon, deeply structured tasks.

\subsection{Agentic Search with LLMs}
With the rapid advancement of LLM reasoning capabilities, recent work has shifted toward agentic search frameworks that interleave multi-step reasoning with tool invocation~\cite{searchr1,research,searcho1}. 
To improve deep reasoning and query formulation, several methods optimize LLMs using reinforcement learning, such as Group Relative Policy Optimization (GRPO), together with verified or learned reward models~\cite{searchr1,research,r1searcher}. 
Other approaches further introduce intermediate summarization or refinement steps to mitigate noisy information and improve task-solving accuracy~\cite{searcho1,simpledeepsearcher,autorefine}.

Despite these advances, most existing agentic search systems rely on unstructured natural language reasoning, which becomes increasingly brittle as reasoning depth grows and task structures become more complex. In addition, the lack of explicit mechanisms for context management also leads to rapid context expansion and information dilution, hindering scalable long-horizon reasoning. 
In contrast, \ours{} addresses these limitations by introducing a structured reasoning protocol that explicitly organizes agent actions, together with a context register that compactly tracks essential reasoning states. This design enables more stable, interpretable, and scalable agentic search over complex, multi-step tasks. 

\section{Method}\label{sec:method}
\begin{figure*}
    \centering
    \includegraphics[width=1\linewidth]{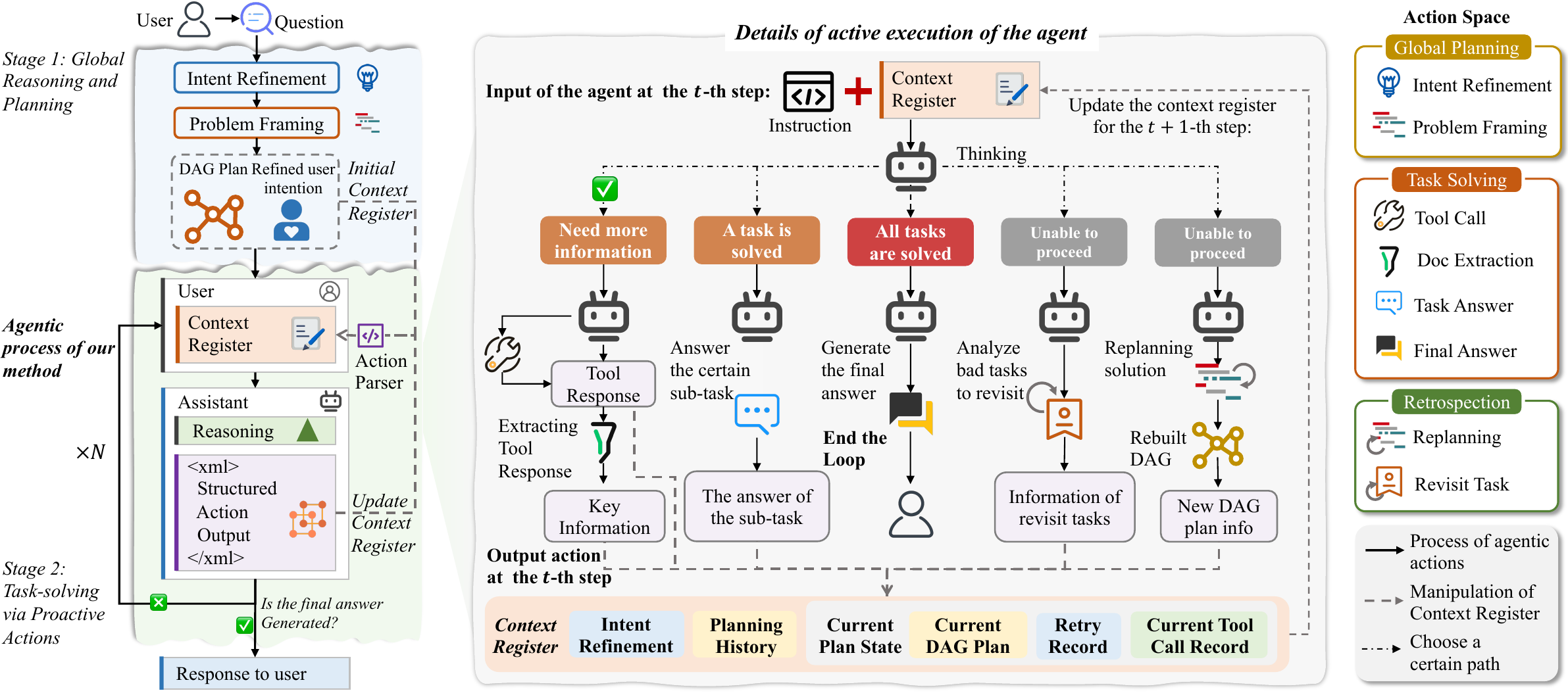}
    \caption{The visualized framework of \ours{}.}
    \label{fig:model}
\end{figure*}
In this section, we demonstrate the framework of our proposed agentic search system, \ours{}, which introduces the structured protocol to organize the reasoning process and the context register to manage reasoning actions and context token usage. Figure~\ref{fig:model} visualizes the overall framework.

\subsection{Structured Protocol}\label{sec:structured_protocol}
To ensure reliable, interpretable, and machine-controllable agentic search, \ours{} defines a structured protocol governing the entire reasoning and interaction process. Unlike conventional agents that produce unconstrained natural-language traces, \ours{} operates through a general and finite set of action units. Each unit is defined by a JSON object, containing its name, description, and standardized executable format, enabling deterministic parsing, tracking, and verification.

We categorize the complete action space into three complementary subspaces: \textit{planning}, \textit{task-solving}, and \textit{retrospection}. Formally, the complete action space is denoted by 
\begin{align}\small
    \mathcal{A} = \mathcal{A}_\textrm{plan} \cup \mathcal{A}_\textrm{sol} \cup \mathcal{A}_\textrm{ret}.
\end{align}
This tripartite decomposition is motivated by long-standing cognitive theories which suggest that intelligent problem solving requires structured processes for problem formulation, goal-directed execution, and metacognitive monitoring~\cite{newell1972human,nelson1990metamemory}. By aligning with this principled organization, our protocol captures the minimal yet sufficient components required to govern coherent and controllable agentic reasoning.

\paragraph{Planning Space} 
The planning space governs the agent’s high-level understanding of user queries and their decomposition into actionable sub-tasks. It includes: (1) \textbf{Intent Refinement} (<intent\_refinement>).
This action clarifies vague, underspecified, or multi-intent user queries by restating a clarified interpretation of the user intent along with any identified constraints. It ensures that the search process begins with a well-defined objective. 
(2) \textbf{Problem-Framing} (<problem\_framing>). This action constructs a graph-structured task decomposition, formalized as a DAG (Directed Acyclic Graph) plan. Each node corresponds to a sub-task, and edges encode dependency relationships. The resulting DAG provides a global structural blueprint, guiding execution order and ensuring coverage of all necessary components.

Together, these planning actions ensure that the reasoning process is grounded in explicit, machine-interpretable structure, rather than relying on loosely organized natural-language reasoning.

\paragraph{Task-solving Space} 
The task-solving space contains actions that execute the plan and interact with external tools to solve tasks. It includes: (1) \textbf{Tool Calling} (<tool\_call>), 
which could invoke external tools, (\eg, retrieval APIs or specialized modules) to gather additional information beyond LLMs' world knowledge. 
(2) \textbf{Document Extraction} (<doc\_extraction>). 
This action analyzes tool responses and extracts key facts valuable to the current task. By condensing retrieved information, this action mitigates uncontrolled context growth from long retrieved documents. 
(3) \textbf{Task Answer} (<task\_answer>). 
When sufficient evidence has been accumulated for one or more sub-tasks in the DAG, the agent uses this action to answer those sub-tasks.
(4) \textbf{Final-Answer} (<final\_answer>). 
After all sub-tasks have been solved, this action produces the global answer to the user’s query. This answer is then surfaced as the system’s final response. 

\paragraph{Retrospection Space} 
The retrospection space equips \ours{} with explicit mechanisms for self-monitoring and correction, enabling robustness against erroneous or unsolvable reasoning paths. It consists of:
(1) \textbf{Revisit-Task} (<revisit\_task>). 
It implements the task-level retrospection. When a previously solved sub-task is suspected to be incorrect, which may potentially block the progress of the current reasoning path, the agent re-executes the sub-task to produce corrected results. 
(2) \textbf{Replanning} (<replanning>). 
This action enables the plan-level retrospection. When the current reasoning trajectory becomes unsolvable due to an insufficient or flawed DAG decomposition, this action triggers reconstruction of an improved DAG plan.

These reflective actions confer a key capability of agentic reasoning: the ability to detect mistakes, adjust plan, and revise reasoning trajectories, rather than rigidly adhering to a faulty path.

\subsection{Context Register}\label{sec:context_register}
A critical challenge in long-horizon agentic search lies in maintaining a coherent, compact, and logical memory of the evolving reasoning process and avoiding the key information being overwhelmed by noise during the reasoning process. The reasoning process of existing methods is usually based on natural language, and the context window will be filled up quickly as the number of reasoning steps increases, which are prone to redundancy, hallucination, and context drift. 

To address this, \ours{} introduces the \textit{context register} module, a structured, slot-based storage module that persistently records essential reasoning state throughout the reasoning lifecycle. 
Formally, the register at step $t$- is represented as:
\begin{align}
    \mathcal{R}^t=\{\mathcal{I},\mathcal{P}_{\textrm{his}},\mathcal{P}_{\textrm{cur}}\}.
\end{align}
Each slot has a clearly defined semantic role:

\noindent(1) Refined intent ($\mathcal{I}$). It stores the refined user intent output by the agentic searcher.

\noindent(2) Planning history ($\mathcal{P}_\textrm{his}$). The archive of ``discarded'' DAG plans: 
\begin{align}
\mathcal{P}_\textrm{his}=\{\mathcal{G}_0, ...\mathcal{G}_{i-1}\},
\end{align}
preventing the agent from re-entering previously invalid planning states.

\noindent (3) Current plan state ($\mathcal{P}_\textrm{cur}$). The active DAG plan together with execution metadata:
\begin{align}
\mathcal{P}_\textrm{cur} = \{\mathcal{G}_i\, \mathcal{V}_\textrm{his}, \mathcal{L}_\textrm{tool}\},
\end{align}
where $\mathcal{V}_\textrm{his}$ records re-executed tasks along with their discarded answers, and $\mathcal{L}_\textrm{tool}$ maintains tool calling logs with condensed key information.

\subsection{Agentic Loop of \ours{}}\label{sec:agent_loop}
Building on the structured protocol and context register, \ours{} operates through a two-stage agentic loop that drives end-to-end task-solving. The first stage, \textit{holistic planning}, interprets user intent and proactively decomposes the task into a structured plan. The second stage, \textit{proactive solving}, 
executes reasoning steps, calls external tools, performs spontaneous self-reflection, and continually updates the context register to drive the solution process toward the final answer.

During holistic planning, \ours{} receives the global instruction and the user query as inputs and executes intent refinement followed by problem framing, formalized as:
\begin{align}
    \mathcal{I} = \mathcal{M}_{\theta}(p_{h}, q) ;\;\;\;\mathcal{G} = \mathcal{M}_{\theta}(p_{h}, q, \mathcal{I}),
\end{align}
where $\mathcal{M}_{\theta}$ denotes the LLM used to build the search agent; $p_{h}$ denotes the prompt of the holistic planning; $\mathcal{I}$ is the refined and structured user intent, which is a JSON data, including attributes of ``refined goal'' and ``constraints''; $\mathcal{G}$ is the structure of the DAG graph containing the decomposed tasks and their relationships. 

Once holistic planning is complete, the context register $\mathcal{R}^{0}$ is initialized from these outputs:
\begin{align}
    \mathcal{R}^0 = \mathcal{C}({\mathcal{I}, \mathcal{G}}).
\end{align}

In the proactive solving stage, the search agent iteratively determines the next action based on the context, and updates the register accordingly: 
\begin{align}
    a^t &= \mathcal{M}_{\theta}(p_{s}, \mathcal{R}^{t-1}, q), a^t \in \mathcal{A}_\textrm{sol} \cup \mathcal{A}_\textrm{ret}; \\
    \mathcal{R}^{t} &= \mathcal{U}(\mathcal{R}^{t-1}, a^t),
\end{align}
where $a^t$ denotes the actions executed in the $t$-th step, $p_{s}$ is the prompt for the proactive solving stage, and $\mathcal{U}$ denotes the update of the context register in the $t$-th step. 
Thanks to the structured protocol, updating the context register requires no additional LLM inference, which is performed purely through deterministic code that parses the model-generated reasoning actions.

\begin{table*}[!ht]
\caption{The overall experimental results of \ours{}. The best and second-best results are in bold and underlined, respectively. ``AVG.'' and ``Impro.'' are short for ``Average'' and ``Improvement''.}
\resizebox{0.99\linewidth}{!}{
\begin{tabular}{llccccccccc}
\toprule
Base Model                 & Method    & \multicolumn{3}{c}{BrowseComp-ZH}                                             & \multicolumn{3}{c}{Bamboogle}                                                                                & \multicolumn{3}{c}{MuSiQue}                                                                                  \\
\cmidrule(lr){3-5}\cmidrule(lr){6-8}\cmidrule(lr){9-11}
                           &           & LJFT & LLMJ & ACC & LJFT                               & LLMJ                               & ACC                                & LJFT                               & LLMJ                               & ACC                                \\
\midrule
                           & No-thinking & 3.11 & 2.42 & 0.69 & 20.80 & 26.40 & 22.40 & 4.20 & 6.00 & 3.00  \\
\multirow{5}{*}{Qwen3-8B}  & Thinking   & 5.88 & 4.15 & 1.73 & 44.80 & 53.60 & 42.40 & 5.80 & 8.60 & 3.40  \\
                           & RAG        & \underline{10.74} & 9.00 & 2.77 & 28.80 & 33.60 & 26.40 & 9.00 & 10.80 & 7.40  \\
                           & ReAct      & 9.34 & \underline{10.69} & \underline{6.23} & \underline{57.60} & \underline{60.00} & \underline{52.00} & \underline{19.40} & \underline{24.60} & \underline{14.60}  \\
                           & Search-R1  & 8.65 & 9.34 & 5.19 & 43.22 & 52.00 & 39.20 & 10.40 & 14.60 & 7.60  \\
\cmidrule(lr){2-11}
                         \rowcolor{blue!10}   &  \ours{}    & \textbf{14.88} & \textbf{13.84} & \textbf{9.00} & \textbf{60.80} & \textbf{67.20} & \textbf{52.80} & \textbf{22.40} & \textbf{28.20} & \textbf{20.40}  \\
\midrule
                           & No-thinking   & 5.19 & 4.15 & 1.38 & 33.60 & 40.80 & 30.40 & 7.80 & 11.20 & 5.80 \\
\multirow{5}{*}{Qwen3-32B} & Thinking       & 10.73 & 10.38 & 4.15 & 58.40 & 62.40 & 54.40 & 10.60 & 14.00 & 7.80 \\
                           & RAG            & 8.30 & 7.61 & 4.15 & 48.00 & 55.20 & 43.20 & 12.20 & 14.00 & 10.00 \\
                           & ReAct          & 14.19 & 15.22 & 7.27 & 55.20 & 61.60 & 48.80 & \underline{22.00} & \underline{26.80} & \underline{15.20} \\
                           & Search-R1      & \underline{15.57} & \underline{15.92} & \underline{9.34} & \underline{56.80} & \underline{66.40} & \underline{53.60} & 17.60 & 22.20 & 11.40 \\
\cmidrule(lr){2-11}
                          \rowcolor{blue!10}  & \ours{}        & \textbf{21.45} & \textbf{21.80} & \textbf{14.53} & \textbf{64.80} & \textbf{69.35} & \textbf{59.20} & \textbf{24.20} & \textbf{31.60} & \textbf{22.00} \\
\midrule
                           &           & \multicolumn{3}{c}{Web Dancer}                                                & \multicolumn{6}{c}{AVG.}   \\
\cmidrule(lr){3-5}\cmidrule(lr){6-11}
                           &           & LJFT & LLMJ & ACC & LJFT   & Impro.  & LLMJ  & Impro.   & ACC   &  Impro.     \\
\midrule
                           & No-thinking  & 21.32 & 21.83 & 14.21 & 12.36 & -60.59\% & 14.16 & -59.08\% & 10.08 & -60.98\% \\
\multirow{5}{*}{Qwen3-8B}  & Thinking     & \underline{40.38} & 38.07 & 26.90 & 24.22 & -22.78\% & 26.11  &  -24.57\% & 16.12  &  -37.56\% \\
                           & RAG          & 39.09 & \underline{43.15} & \underline{30.96} & 21.91 & -30.14\% & 24.14  & -30.26\% & 16.88 & -34.62\% \\
                           & ReAct        & 39.09 & \underline{43.15} & 30.46 & \underline{31.36} & - & \underline{34.61} & - & \underline{25.82}  & - \\
                           & Search-R1    & 36.55 & 39.59 & 24.87 & 24.71 & -21.22\% & 28.88 & -16.54\% & 19.22  & -25.58\% \\
\cmidrule(lr){2-11}
                          \rowcolor{blue!10}  & \ours{}      & \textbf{42.13} & \textbf{47.72} & \textbf{29.44} & \textbf{35.05} & \textbf{+11.78\%} & \textbf{39.24} & \textbf{+26.39\%} & \textbf{27.91} & \textbf{+8.09\%} \\
\midrule
                           & No-thinking  & 31.98 & 33.50 & 26.90 & 19.64 & -41.86\% & 22.41 & -42.45\% & 16.12 & -38.76\% \\
\multirow{5}{*}{Qwen3-32B} & Thinking   & 39.59 & 43.65 & 32.99 & 29.83 & -11.71\% & 32.61 & -16.27\% & 24.84 & -5.66\% \\
                           & RAG        & \underline{46.19} & 49.75 & \underline{35.53} & 28.67 & -15.14\% & 36.74 & -5.66\% & 23.22 & -11.80\% \\
                           & ReAct      & 38.07 & 42.13 & 27.92 & 32.37 & -4.21\% & 36.44 & -6.44\% & 24.80 & -5.81\% \\
                           & Search-R1  & 45.18 & \underline{51.27} & 30.96 & \underline{33.79} & - & \underline{38.95} & - & \underline{26.33} & - \\
\cmidrule(lr){2-11}
                          \rowcolor{blue!10}  & \ours{}    & \textbf{49.24} & \textbf{54.82} & \textbf{31.98} & \textbf{39.92} & \textbf{+18.16\%} & \textbf{44.39} & \textbf{+13.99\%} & \textbf{31.93} & \textbf{+21.28\%} \\
\bottomrule
\end{tabular}
}
\label{tab:overall}
\end{table*}

\subsection{Optimization of \ours{}}\label{sec_sft}
Experiments on the Qwen3-series models demonstrate that our agentic search framework achieves strong performance without any additional training. It substantially enhances deep search capabilities while preserving the foundational abilities of the underlying LLMs.

To further ensure reliable structured-action generation on medium-size models and to enable a fair comparison with training-based agentic search systems, we additionally apply Rejection Sampling Fine-Tuning (RFT) to optimize student models. Specifically, we collect high-quality reasoning trajectories from stronger LLMs and retain only those samples that both (1) arrive at correct final answers and (2) follow the required structured output schema. These filtered trajectories serve as our training corpus.
For each query $q$, its reasoning trajectory consists of $T$ state-action pairs:
\begin{align}
    \mathcal{T}(q) &= \{(s^1, a^1), ..., (s^T, a^T)\},\\
    a^1, a^2&\in\mathcal{A}_\textrm{plan}, a^T=a^{fa},\nonumber \\
    a^{3:T-1}&\in\mathcal{A}_\textrm{sol}\cup\mathcal{A}_\textrm{ret}-\{a^{fa}\},\nonumber
\end{align}
where $s^i$ denotes the state (LLM inputs) at the $t$-th step, including the instruction, user question, and the context register. $a^{fa}$ is the final answer action.
\begin{align}
   \mathcal{L} = -\sum_{t=1}^{T}\sum_{i=1}^{|a^t|}\log P_\theta(a^t_i | a^t_{1:i-1}, s^t).
\end{align}

\section{Experiment}

\begin{table*}[]
\centering
\caption{Overall experimental results on fine-tuned agentic search baselines. The best and second-best results are in bold and underlined, respectively. ``AVG.'' and ``Impro.'' are short for ``Average'' and ``Improvement''.}
\resizebox{0.99\linewidth}{!}{
\begin{tabular}{lccccccccc}
\toprule
\multicolumn{1}{c}{\multirow{2}{*}{Models}} & \multicolumn{3}{c}{BrowseComp-ZH}                                             & \multicolumn{3}{c}{Bamboogle}                                                                                & \multicolumn{3}{c}{MuSiQue}                                                                                  \\
\cmidrule(lr){2-4}\cmidrule(lr){5-7}\cmidrule(lr){8-10}
                        & LJFT & LLMJ & ACC & LJFT                               & LLMJ                               & ACC                                & LJFT                               & LLMJ                               & ACC                                \\

\midrule
Search-R1                                   & 3.11                     & 3.46                     & 1.38                    & 52.80          & {56.80}          & {47.20}          & {20.20}          & {27.20}    & {16.80}          \\
AutoRefine                                   & 5.19                    &        5.19          & 3.11                   &      48.00                     &        53.60                  &   42.40                          &    19.80                      &       25.20                     &    16.00                    \\
R1-Searcher                                 & 3.46                     & 3.81                     & 0.35                    & {{56.80}}    &  {56.80}          & {48.80}          & \underline{22.40}    & {\textbf{28.80}} & {\underline{18.80}} \\
SimpleDeepSearcher                          & 7.27                     & 6.92                     & \underline{2.77}              & {\textbf{61.60}} & {\textbf{65.60}} & {\textbf{57.60}} & {21.20}          & {25.80}          & {{18.20}}    \\
\midrule
\rowcolor{blue!10}  \ours{}                                       & \textbf{9.69}            & \textbf{9.34}            & \textbf{3.46}           & {\textbf{61.60}} & {\textbf{65.60}} & {\underline{56.80}} & \textbf{23.60}          & \underline{27.20}          & {\textbf{19.20}}   \\
\midrule
                                            & \multicolumn{3}{c}{Web Dancer}                                                & \multicolumn{6}{c}{AVG.}  \\
\cmidrule(lr){2-4}\cmidrule(lr){5-10}
                                            & LJFT & LLMJ & ACC & LJFT                               & Impro.                             & LLMJ                               & Impro.                             & ACC                                & Impro.                             \\
\midrule
Search-R1                                   & 38.07                    & 44.67                    & 28.43                   & 28.55                              & -15.26\%                           & 33.03                              & -9.93\%                            & 23.45                              & -16.67\%                           \\
AutoRefine                                   & 39.09                    &       44.16             & 26.90                   &      28.02  &	-16.82\%    &	32.04    &	-12.85\%	&  22.10     & 	-21.46\%                 \\
R1-Searcher                                 & 36.55                    & 41.62                    & 26.90                   & 29.80                              & -11.52\%                           & 32.76                              & -10.89\%                           & 23.71                              & -15.74\%                           \\
SimpleDeepSearcher                          & \underline{44.67}              & \underline{48.73}              & \underline{34.01}             & \underline{33.68}                        &                -                   & \underline{36.76}                        &                  -                 & \underline{28.14}                        &   -                                \\
\midrule
\rowcolor{blue!10}  \ours{}                                       & \textbf{47.72}           & \textbf{54.31}           & \textbf{38.07}          & \textbf{35.65}                     & \textbf{+5.84\%}                             & \textbf{39.11}                     & \textbf{++6.40\%}                             & \textbf{29.38}                     & \textbf{++4.40\%}         \\
\bottomrule
\end{tabular}
}
\label{tab:sft_overall}
\end{table*}

\paragraph{Datasets and Baselines}
We evaluate the prompting-based version of \ours{} using the Qwen3-8B and Qwen3-32B models~\cite{qwen3} on several challenging multi-hop question-answering (QA) benchmarks, including BrowseComp-ZH~\cite{browsecomp-zh}, WebDancer~\cite{webdancer}, Bamboogle~\cite{bamboogle}, and MuSiQue~\cite{musique}. For MuSiQue, we randomly sample 500 examples for evaluation due to its large scale.
We compare \ours{} with a wide range of LLM-based information-seeking frameworks:
(1) basic ``No-thinking'' and ``Thinking'' modes without external retrieval;
(2) retrieval-augmented generation (RAG);
(3) natural-language-based (NL-based) agentic search pipelines, including ReAct~\cite{yao2023react} and Search-R1~\cite{searchr1} (without training).

For the optimized version of \ours{}, we fine-tune a Qwen2.5-7B-base model to ensure a fair comparison with existing agentic search systems, including Search-R1~\cite{searchr1}, AutoRefine~\cite{autorefine}, R1-Searcher~\cite{r1searcher}, and SimpleDeepSearcher~\cite{simpledeepsearcher}.
Concretely, we use DeepSeek-V3.1~\cite{deepseekv3} as the teacher model to collect reasoning trajectories from Asearcher~\cite{asearcher} and MiroRL-GenQA~\cite{mirorl} datasets, yielding 551 training QA instances with 7,944 state–action pairs after quality filtering. We set the learning rate as 1e-5, the epoch as 5, and the warm up ratio as 0.1 to optimize our model. 

We use a commercial search engine for BrowseComp-ZH and WebDancer, and a local E5 retriever~\cite{e5} over the 2018 Wikipedia dump~\cite{dpr}, following~\citet{searchr1} for other datasets. The agent context length is set to 32,000 tokens, and the top-3 documents are retrieved per query. All models share the same retrieval settings.

\paragraph{Evaluation Metrics}
To assess answer correctness, we report both rule-based accuracy (ACC)~\cite{searcho1} and LLM-as-a-Judge~\cite{llmasajudge} scores for improved semantic evaluation.
We implement two variants of LLM-as-a-Judge:
(1) a fine-tuned reward model following the xVerify framework~\cite{xVerify}, trained on proprietary curated data and achieving 98.2\% accuracy on a commercial human-annotated benchmark. We denote this metric as LJFT;
(2) a zero-shot judge model following~\cite{simpledeepsearcher}, using DeepSeek-V3.1-Terminus~\cite{deepseekv3} as the reward model, denoted as LLMJ.

\begin{figure*}[!h]
    \centering
    \includegraphics[width=1\linewidth]{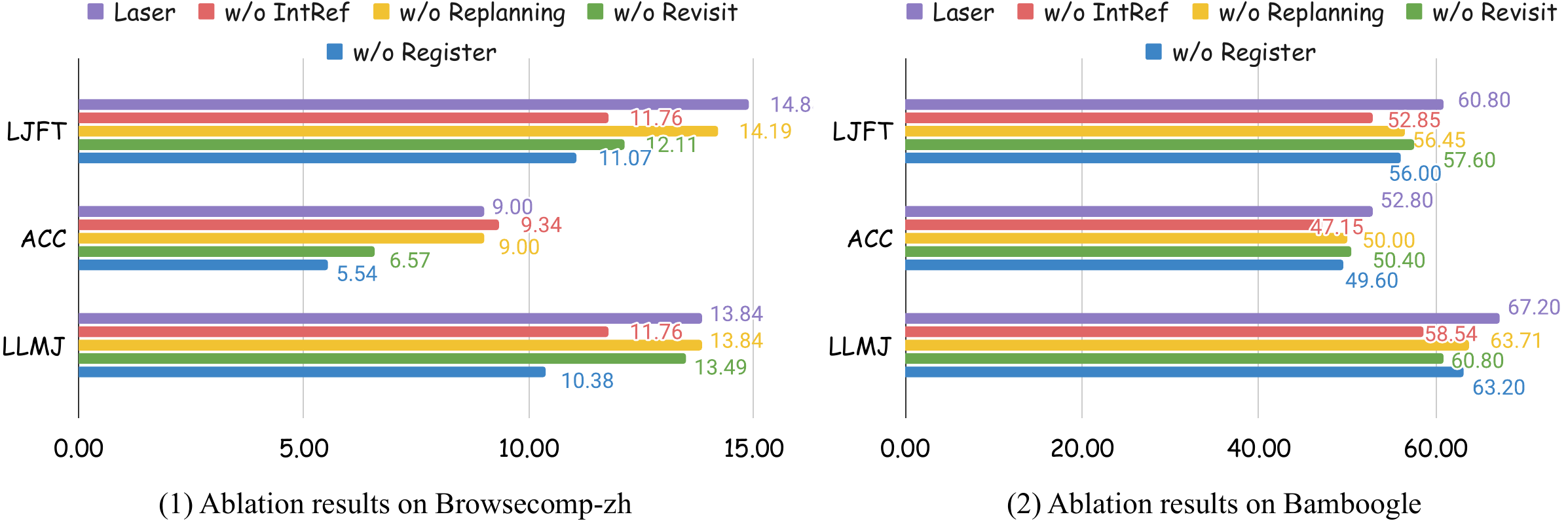}
    \caption{Ablation results.}
    \label{fig:abl}
\end{figure*}
\begin{figure}[!ht]
    \centering
    \includegraphics[width=1\linewidth]{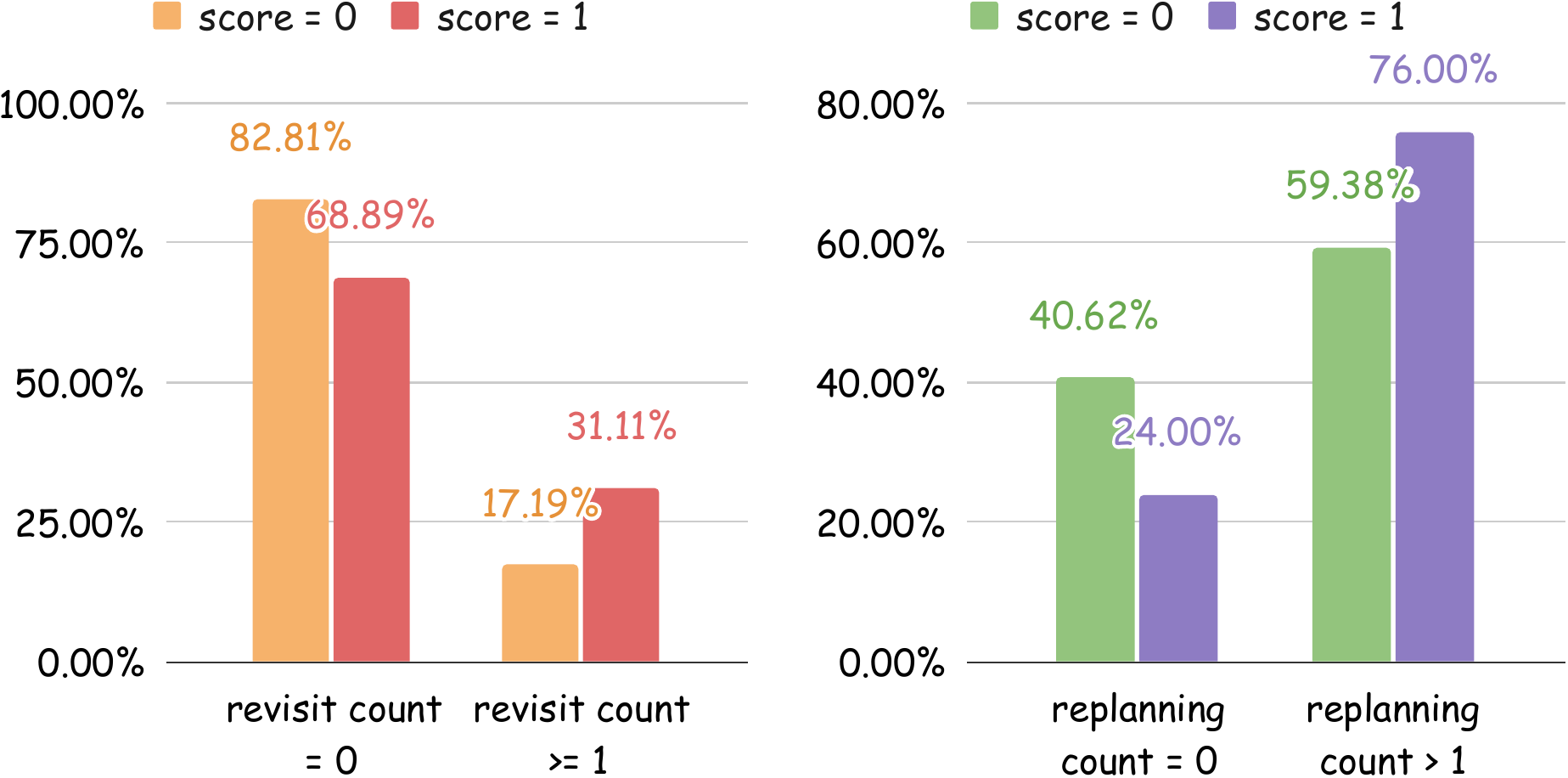}
    \caption{Score distributions between different retrospection scenarios.}
    \label{fig:retrospection}
\end{figure}
\subsection{Overall Experiments}
The overall results are summarized in Tables~\ref{tab:overall} and~\ref{tab:sft_overall}. Across nearly all datasets and evaluation metrics, \ours{} demonstrates substantial improvements over all baselines in both prompting and fine-tuning settings. The consistently superior average performance highlights the effectiveness of our approach: structured logical reasoning combined with register-based context management enables markedly stronger agentic search capabilities than prior natural-language–based pipelines.

We also observe that with the help of external knowledge, the LLM performs significantly better than close-book settings, especially for no-thinking version, which proves the valuable of web search for reliable and accurate LLM-based information seeking systems. Moreover, the search agent models also generally outperform RAG methods, which also highlight the importance of agentic search systems in dealing with long-horizon and complex user tasks.

\subsection{Ablation Analyses}
\begin{figure*}[!ht]
    \centering
    \includegraphics[width=1\linewidth]{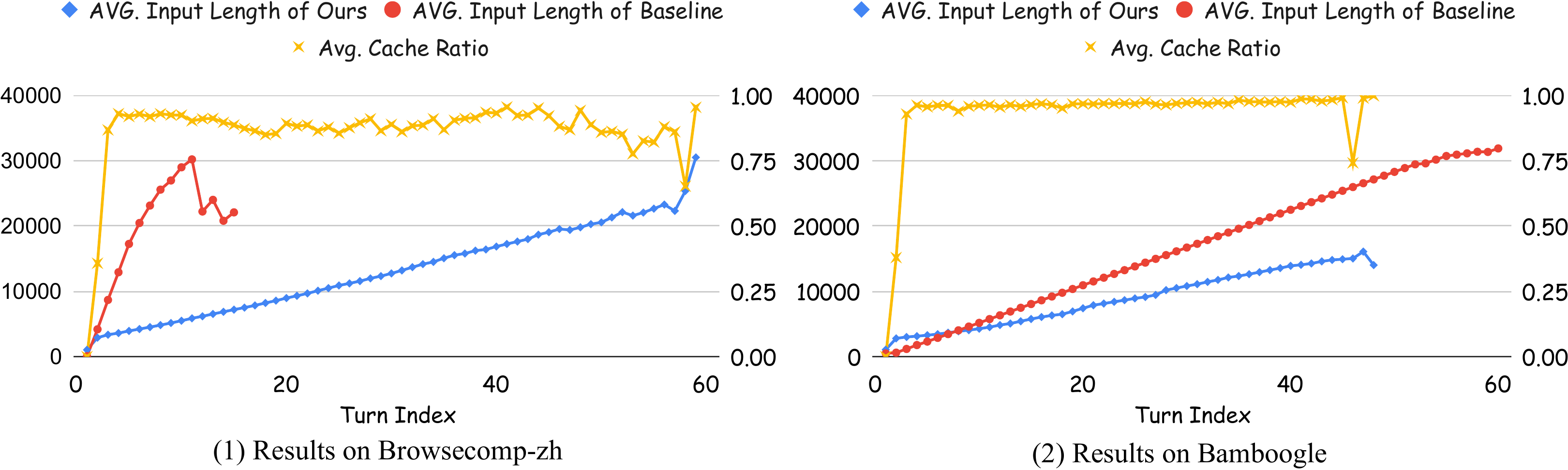}
    \caption{Visualization of growth trends of context length and cache ratio.}
    \label{fig:context}
\end{figure*}
\begin{figure*}[!ht]
    \centering
    \includegraphics[width=1\linewidth]{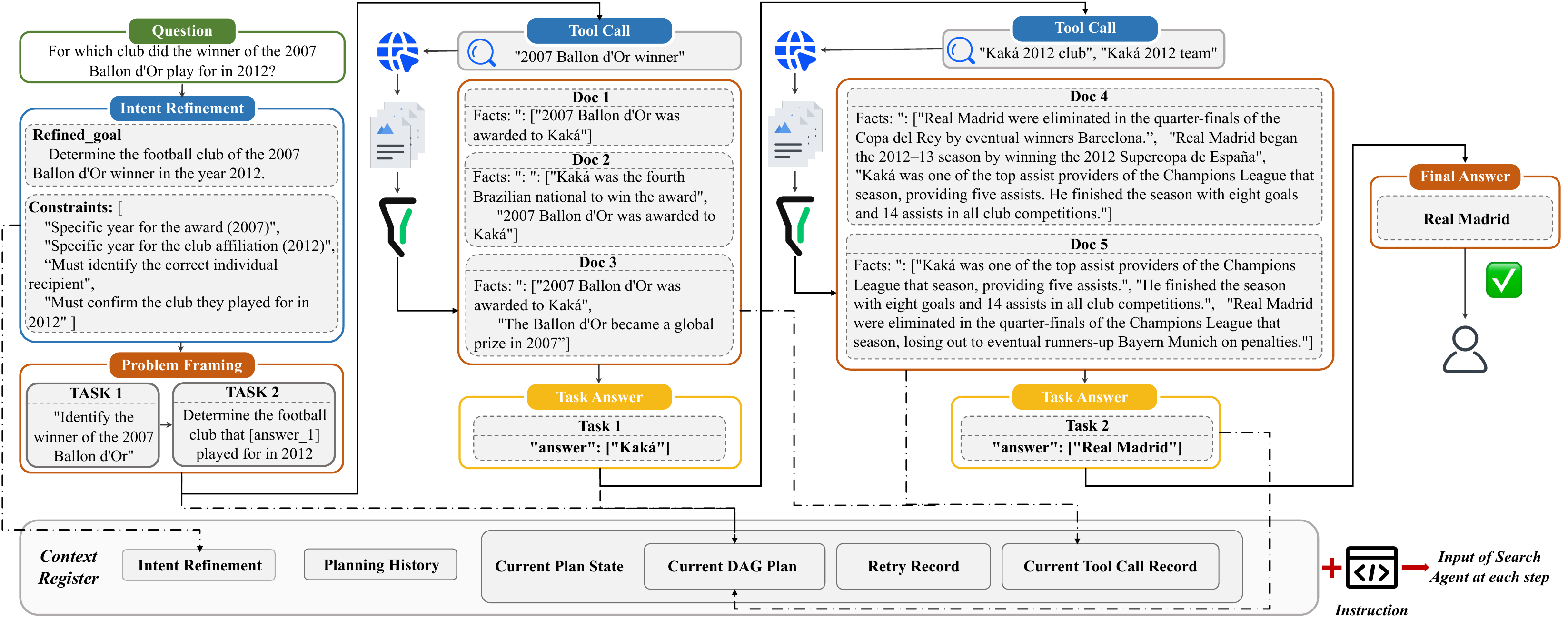}
    \caption{The visualization of a case of \ours{}.}
    \label{fig:case}
\end{figure*}

To assess the contribution of each core component in our framework, we conduct a series of ablation experiments targeting the major design modules.

First, to examine the role of intent refinement, we remove this action and construct a variant denoted as w/o IntRef. The results (second row in Figure~\ref{fig:abl}) show a clear performance drop, highlighting the importance of explicit intent interpretation. Natural-language user queries often contain ambiguity or incomplete specifications; refining the intent and extracting explicit constraints enables the agent to establish a clearer objective, which in turn facilitates stable multi-step reasoning.

Next, to evaluate the contribution of the retrospection space, we separately remove the replanning and revisit-task actions, producing the variants w/o Replanning and w/o Revisit. As shown in the third and fourth rows of Figure~\ref{fig:abl}, performance declines noticeably in both settings. This demonstrates the necessity of incorporating structured retrospection: complex multi-hop tasks rarely follow a perfect reasoning trajectory on the first attempt, and the ability to backtrack, repair errors, and reconstruct plans is crucial for robust agentic search systems.

Finally, we evaluate the effect of the context register by disabling it entirely, yielding the variant w/o Register. The substantial degradation in the last row of Figure~\ref{fig:abl} confirms its importance. The context register filters noise, preserves only essential state variables, and prevents the context window from being overwhelmed by verbose natural-language traces. This enables the agent to sustain longer reasoning trajectories while maintaining stable context quality and controllability.

\subsection{Retrospection Analyses}

To further examine the contribution of retrospection actions, we categorize samples according to whether their reasoning trajectories contain revisit-task or replanning actions, and visualize the distribution of correct and incorrect answers in Figure~\ref{fig:retrospection}.

The results reveal that samples with correct final answers exhibit substantially higher execution rates of retrospection actions than those with incorrect answers. This trend indicates that retrospection, whether at the task level or the plan level, plays an essential role in guiding the agent out of faulty reasoning paths and correcting intermediate errors.

In addition, we observe that the replanning action is executed more frequently than the revisit-task action. This suggests that when encountering impasses, the agent is more inclined to restructure the global DAG plan rather than re-solving individual sub-tasks. Such behavior implies that for many complex multi-hop queries, inadequacies in the initial task decomposition may be a more common bottleneck than isolated sub-task errors, making global plan reconstruction a more effective corrective strategy.

\subsection{Context Analyses}
We further analyze the evolution of context length across reasoning turns to illustrate the impact of the proposed context register. Specifically, we compare \ours{} with ReAct and additionally report the token cache ratio, defined as the proportion of prefix tokens shared between two consecutive input sequences. Figure~\ref{fig:context} visualizes these results.

As shown, \ours{} exhibits a markedly slower context growth rate than ReAct. In ReAct, natural-language traces accumulate rapidly, and the context window is typically saturated within 20 steps, preventing further reasoning. In contrast, \ours{} maintains compact and noise-reduced state representations through the context register, enabling more reasoning turns while preserving clarity and reducing drift. This allows the system to perform deeper, higher-quality multi-hop reasoning. 
Moreover, \ours{} achieves consistently high token cache ratios, indicating great reuse of the prefix across turns. This property not only stabilizes the reasoning trajectory but also reduces computational overhead by enabling more efficient token caching, thereby striking a favorable balance between effectiveness and efficiency.

We notice that the data points  at the end of the curve are usually somewhat abnormal (the curve connection with the previous data points is not smooth). It is because the number of reasoning trajectories with a considerable number of rounds rounds is small, and thus the mean distribution is easily affected by outliers and increased variance, rather than reflecting a systematic value drop. 

\subsection{Case Studies}
We also provide an example in Figure~\ref{fig:case} to visualize the agentic process of \ours{}. Due to the limited space, we select a relatively easy case (a two-hop question) in the Bamboogle dataset . The case presents that once receiving the user question, \ours{} will execute intent refinement and problem-framing actions to understand and formulate the solving task in the global perspective, which is used to build the initial content of the context register. Then, the search agent executes the specific tool (search engine) to retrieve relevant information to solve the current sub-task. The retrieved documents will be extracted key facts related to the user question to remove the massive noise. When the agent believes the acquired information is enough to solve the current task, it will generate the task answer for the current task and then transfer to subsequent sub-tasks. All these action outputs are formulated following our action schema and can be parsed by Python code. The parsed results will be used to update the context register, serving as the LLM's input at each reasoning step, therefore supporting sufficient rounds of reasoning and maintaining a slow increase in the length of the context.

\section{Conclusion}
This work addresses two fundamental limitations of existing agentic search systems: unstable reasoning induced by unstructured natural-language trajectories and performance degradation caused by uncontrolled context growth. We propose \ours{}, a general and principled framework that stabilizes agentic search through a symbolic action protocol with explicit semantics and a compact context register that preserves essential reasoning state. By structuring agent behaviors across planning, task-solving, and retrospection, \ours{} enables logically coherent, long-horizon reasoning while remaining context-efficient. Extensive experiments on challenging multi-hop question answering benchmarks demonstrate consistent and significant improvements over prior agentic search approaches under both prompting and fine-tuning regimes. Together, these results suggest that explicitly structured reasoning and disciplined context management are key to building robust and scalable agentic search systems, offering a promising direction for future reasoning-centric agents.



\bibliography{custom}

@String{Computer = "{IEEE} Computer" }

@article{RichRAG,
  author       = {Shuting Wang and
                  Xin Yu and
                  Mang Wang and
                  Weipeng Chen and
                  Yutao Zhu and
                  Zhicheng Dou},
  title        = {RichRAG: Crafting Rich Responses for Multi-faceted Queries in Retrieval-Augmented
                  Generation},
  journal      = {CoRR},
  volume       = {abs/2406.12566},
  year         = {2024},
  url          = {https://doi.org/10.48550/arXiv.2406.12566},
  doi          = {10.48550/ARXIV.2406.12566},
  eprinttype    = {arXiv},
  eprint       = {2406.12566},
  bibsource    = {dblp computer science bibliography, https://dblp.org}
}

@book{newell1972human,
  title={Human problem solving},
  author={Newell, Allen and Simon, Herbert Alexander and others},
  volume={104},
  number={9},
  year={1972},
  publisher={Prentice-hall Englewood Cliffs, NJ}
}

@incollection{nelson1990metamemory,
  title={Metamemory: A theoretical framework and new findings},
  author={Nelson, Thomas O},
  booktitle={Psychology of learning and motivation},
  volume={26},
  pages={125--173},
  year={1990},
  publisher={Elsevier}
}

@article{atlas,
author = {Izacard, Gautier and Lewis, Patrick and Lomeli, Maria and Hosseini, Lucas and Petroni, Fabio and Schick, Timo and Dwivedi-Yu, Jane and Joulin, Armand and Riedel, Sebastian and Grave, Edouard},
title = {Atlas: few-shot learning with retrieval augmented language models},
year = {2024},
issue_date = {January 2023},
publisher = {JMLR.org},
volume = {24},
number = {1},
issn = {1532-4435},
journal = {J. Mach. Learn. Res.},
month = {mar},
articleno = {251},
numpages = {43}
}

@InProceedings{retro,
  title = 	 {Improving Language Models by Retrieving from Trillions of Tokens},
  author =       {Borgeaud, Sebastian and Mensch, Arthur and Hoffmann, Jordan and Cai, Trevor and Rutherford, Eliza and Millican, Katie and Van Den Driessche, George Bm and Lespiau, Jean-Baptiste and Damoc, Bogdan and Clark, Aidan and De Las Casas, Diego and Guy, Aurelia and Menick, Jacob and Ring, Roman and Hennigan, Tom and Huang, Saffron and Maggiore, Loren and Jones, Chris and Cassirer, Albin and Brock, Andy and Paganini, Michela and Irving, Geoffrey and Vinyals, Oriol and Osindero, Simon and Simonyan, Karen and Rae, Jack and Elsen, Erich and Sifre, Laurent},
  booktitle = 	 {Proceedings of the 39th International Conference on Machine Learning},
  pages = 	 {2206--2240},
  year = 	 {2022},
  editor = 	 {Chaudhuri, Kamalika and Jegelka, Stefanie and Song, Le and Szepesvari, Csaba and Niu, Gang and Sabato, Sivan},
  volume = 	 {162},
  series = 	 {Proceedings of Machine Learning Research},
  month = 	 {17--23 Jul},
  publisher =    {PMLR},
  url = 	 {https://proceedings.mlr.press/v162/borgeaud22a.html}
}

@inproceedings{retroplus,
    title = "Shall We Pretrain Autoregressive Language Models with Retrieval? A Comprehensive Study",
    author = "Wang, Boxin  and
      Ping, Wei  and
      Xu, Peng  and
      McAfee, Lawrence  and
      Liu, Zihan  and
      Shoeybi, Mohammad  and
      Dong, Yi  and
      Kuchaiev, Oleksii  and
      Li, Bo  and
      Xiao, Chaowei  and
      Anandkumar, Anima  and
      Catanzaro, Bryan",
    editor = "Bouamor, Houda  and
      Pino, Juan  and
      Bali, Kalika",
    booktitle = "Proceedings of the 2023 Conference on Empirical Methods in Natural Language Processing",
    month = dec,
    year = "2023",
    address = "Singapore",
    publisher = "Association for Computational Linguistics",
    url = "https://aclanthology.org/2023.emnlp-main.482",
    doi = "10.18653/v1/2023.emnlp-main.482",
    pages = "7763--7786",
}

@article{realm,
  title={GAR-meets-RAG Paradigm for Zero-Shot Information Retrieval},
  author={Daman Arora and Anush Kini and Sayak Ray Chowdhury and Nagarajan Natarajan and Gaurav Sinha and Amit Sharma},
  journal={ArXiv},
  year={2023},
  volume={abs/2310.20158},
  url={https://api.semanticscholar.org/CorpusID:264817661}
}

@inproceedings{hindsight,
  author       = {Ashwin Paranjape and
                  Omar Khattab and
                  Christopher Potts and
                  Matei Zaharia and
                  Christopher D. Manning},
  title        = {Hindsight: Posterior-guided training of retrievers for improved open-ended
                  generation},
  booktitle    = {The Tenth International Conference on Learning Representations, {ICLR}
                  2022, Virtual Event, April 25-29, 2022},
  publisher    = {OpenReview.net},
  year         = {2022},
  url          = {https://openreview.net/forum?id=Vr\_BTpw3wz},
  bibsource    = {dblp computer science bibliography, https://dblp.org}
}

@inproceedings{rag,
author = {Lewis, Patrick and Perez, Ethan and Piktus, Aleksandra and Petroni, Fabio and Karpukhin, Vladimir and Goyal, Naman and K\"{u}ttler, Heinrich and Lewis, Mike and Yih, Wen-tau and Rockt\"{a}schel, Tim and Riedel, Sebastian and Kiela, Douwe},
title = {Retrieval-augmented generation for knowledge-intensive NLP tasks},
year = {2020},
isbn = {9781713829546},
publisher = {Curran Associates Inc.},
address = {Red Hook, NY, USA},
booktitle = {Proceedings of the 34th International Conference on Neural Information Processing Systems},
articleno = {793},
numpages = {16},
location = {<conf-loc>, <city>Vancouver</city>, <state>BC</state>, <country>Canada</country>, </conf-loc>},
series = {NIPS '20}
}

@article{radit,
  title={RA-DIT: Retrieval-Augmented Dual Instruction Tuning},
  author={Xi Victoria Lin and Xilun Chen and Mingda Chen and Weijia Shi and Maria Lomeli and Rich James and Pedro Rodriguez and Jacob Kahn and Gergely Szilvasy and Mike Lewis and Luke Zettlemoyer and Scott Yih},
  journal={ArXiv},
  year={2023},
  volume={abs/2310.01352},
  url={https://api.semanticscholar.org/CorpusID:263605962}
}

@inproceedings{self-rag,
author={Asai, Akari and Wu, Zeqiu and Wang, Yizhong and Sil, Avirup and Hajishirzi, Hannaneh},
title={Self-{RAG}: Learning to Retrieve, Generate, and Critique through Self-Reflection},
booktitle={The Twelfth International Conference on Learning Representations},
year={2024},
url={https://openreview.net/forum?id=hSyW5go0v8}
}

@misc{replug,
      title={REPLUG: Retrieval-Augmented Black-Box Language Models}, 
      author={Weijia Shi and Sewon Min and Michihiro Yasunaga and Minjoon Seo and Rich James and Mike Lewis and Luke Zettlemoyer and Wen-tau Yih},
      year={2023},
      eprint={2301.12652},
      archivePrefix={arXiv},
      primaryClass={cs.CL}
}

@inproceedings{ARR,
    title = "Augmentation-Adapted Retriever Improves Generalization of Language Models as Generic Plug-In",
    author = "Yu, Zichun  and
      Xiong, Chenyan  and
      Yu, Shi  and
      Liu, Zhiyuan",
    editor = "Rogers, Anna  and
      Boyd-Graber, Jordan  and
      Okazaki, Naoaki",
    booktitle = "Proceedings of the 61st Annual Meeting of the Association for Computational Linguistics (Volume 1: Long Papers)",
    month = jul,
    year = "2023",
    address = "Toronto, Canada",
    publisher = "Association for Computational Linguistics",
    url = "https://aclanthology.org/2023.acl-long.136",
    doi = "10.18653/v1/2023.acl-long.136",
    pages = "2421--2436",
}

@misc{zhengbao_filter,
      title={Learning to Filter Context for Retrieval-Augmented Generation}, 
      author={Zhiruo Wang and Jun Araki and Zhengbao Jiang and Md Rizwan Parvez and Graham Neubig},
      year={2023},
      eprint={2311.08377},
      archivePrefix={arXiv},
      primaryClass={cs.CL}
}

@misc{xiong_filter,
      title={Say More with Less: Understanding Prompt Learning Behaviors through Gist Compression}, 
      author={Xinze Li and Zhenghao Liu and Chenyan Xiong and Shi Yu and Yukun Yan and Shuo Wang and Ge Yu},
      year={2024},
      eprint={2402.16058},
      archivePrefix={arXiv},
      primaryClass={cs.CL}
}

@misc{comp1,
      title={RECOMP: Improving Retrieval-Augmented LMs with Compression and Selective Augmentation}, 
      author={Fangyuan Xu and Weijia Shi and Eunsol Choi},
      year={2023},
      eprint={2310.04408},
      archivePrefix={arXiv},
      primaryClass={cs.CL}
}

@misc{bgm,
      title={Bridging the Preference Gap between Retrievers and LLMs}, 
      author={Zixuan Ke and Weize Kong and Cheng Li and Mingyang Zhang and Qiaozhu Mei and Michael Bendersky},
      year={2024},
      eprint={2401.06954},
      archivePrefix={arXiv},
      primaryClass={cs.CL}
}

@misc{ragsurvey,
      title={Retrieval-Augmented Generation for Large Language Models: A Survey}, 
      author={Yunfan Gao and Yun Xiong and Xinyu Gao and Kangxiang Jia and Jinliu Pan and Yuxi Bi and Yi Dai and Jiawei Sun and Meng Wang and Haofen Wang},
      year={2024},
      eprint={2312.10997},
      archivePrefix={arXiv},
      primaryClass={cs.CL}
}

@misc{chan2024rqrag,
      title={RQ-RAG: Learning to Refine Queries for Retrieval Augmented Generation}, 
      author={Chi-Min Chan and Chunpu Xu and Ruibin Yuan and Hongyin Luo and Wei Xue and Yike Guo and Jie Fu},
      year={2024},
      eprint={2404.00610},
      archivePrefix={arXiv},
      primaryClass={cs.CL}
}

@misc{wang2024selfdc,
      title={Self-DC: When to retrieve and When to generate? Self Divide-and-Conquer for Compositional Unknown Questions}, 
      author={Hongru Wang and Boyang Xue and Baohang Zhou and Tianhua Zhang and Cunxiang Wang and Guanhua Chen and Huimin Wang and Kam-fai Wong},
      year={2024},
      eprint={2402.13514},
      archivePrefix={arXiv},
      primaryClass={cs.CL}
}

@misc{khot2023decomposed,
      title={Decomposed Prompting: A Modular Approach for Solving Complex Tasks}, 
      author={Tushar Khot and Harsh Trivedi and Matthew Finlayson and Yao Fu and Kyle Richardson and Peter Clark and Ashish Sabharwal},
      year={2023},
      eprint={2210.02406},
      archivePrefix={arXiv},
      primaryClass={cs.CL}
}

@misc{yao2023react,
      title={ReAct: Synergizing Reasoning and Acting in Language Models}, 
      author={Shunyu Yao and Jeffrey Zhao and Dian Yu and Nan Du and Izhak Shafran and Karthik Narasimhan and Yuan Cao},
      year={2023},
      eprint={2210.03629},
      archivePrefix={arXiv},
      primaryClass={cs.CL}
}

@misc{searchchain,
      title={Search-in-the-Chain: Interactively Enhancing Large Language Models with Search for Knowledge-intensive Tasks}, 
      author={Shicheng Xu and Liang Pang and Huawei Shen and Xueqi Cheng and Tat-Seng Chua},
      year={2024},
      eprint={2304.14732},
      archivePrefix={arXiv},
      primaryClass={cs.CL},
      url={https://arxiv.org/abs/2304.14732}, 
}

@inproceedings{searchr1,
title={Search-R1: Training {LLM}s to Reason and Leverage Search Engines with Reinforcement Learning},
author={Bowen Jin and Hansi Zeng and Zhenrui Yue and Jinsung Yoon and Sercan O Arik and Dong Wang and Hamed Zamani and Jiawei Han},
booktitle={Second Conference on Language Modeling},
year={2025},
url={https://openreview.net/forum?id=Rwhi91ideu}
}

@misc{r1searcher,
      title={R1-Searcher: Incentivizing the Search Capability in LLMs via Reinforcement Learning}, 
      author={Huatong Song and Jinhao Jiang and Yingqian Min and Jie Chen and Zhipeng Chen and Wayne Xin Zhao and Lei Fang and Ji-Rong Wen},
      year={2025},
      eprint={2503.05592},
      archivePrefix={arXiv},
      primaryClass={cs.AI},
      url={https://arxiv.org/abs/2503.05592}, 
}

@inproceedings{simpledeepsearcher,
    title = "{S}imple{D}eep{S}earcher: Deep Information Seeking via Web-Powered Reasoning Trajectory Synthesis",
    author = "Sun, Shuang  and
      Song, Huatong  and
      Wang, Yuhao  and
      Ren, Ruiyang  and
      Jiang, Jinhao  and
      Zhang, Junjie  and
      Bai, Fei  and
      Deng, Jia  and
      Zhao, Xin  and
      Liu, Zheng  and
      Fang, Lei  and
      Wang, Zhongyuan  and
      Wen, Ji-Rong",
    editor = "Christodoulopoulos, Christos  and
      Chakraborty, Tanmoy  and
      Rose, Carolyn  and
      Peng, Violet",
    booktitle = "Findings of the Association for Computational Linguistics: EMNLP 2025",
    month = nov,
    year = "2025",
    address = "Suzhou, China",
    publisher = "Association for Computational Linguistics",
    url = "https://aclanthology.org/2025.findings-emnlp.739/",
    doi = "10.18653/v1/2025.findings-emnlp.739",
    pages = "13705--13720",
    ISBN = "979-8-89176-335-7"
}

@inproceedings{searcho1,
    title = "Search-o1: Agentic Search-Enhanced Large Reasoning Models",
    author = "Li, Xiaoxi  and
      Dong, Guanting  and
      Jin, Jiajie  and
      Zhang, Yuyao  and
      Zhou, Yujia  and
      Zhu, Yutao  and
      Zhang, Peitian  and
      Dou, Zhicheng",
    editor = "Christodoulopoulos, Christos  and
      Chakraborty, Tanmoy  and
      Rose, Carolyn  and
      Peng, Violet",
    booktitle = "Proceedings of the 2025 Conference on Empirical Methods in Natural Language Processing",
    month = nov,
    year = "2025",
    address = "Suzhou, China",
    publisher = "Association for Computational Linguistics",
    url = "https://aclanthology.org/2025.emnlp-main.276/",
    doi = "10.18653/v1/2025.emnlp-main.276",
    pages = "5420--5438",
    ISBN = "979-8-89176-332-6"
}

@misc{research,
      title={ReSearch: Learning to Reason with Search for LLMs via Reinforcement Learning}, 
      author={Mingyang Chen and Linzhuang Sun and Tianpeng Li and Haoze Sun and Yijie Zhou and Chenzheng Zhu and Haofen Wang and Jeff Z. Pan and Wen Zhang and Huajun Chen and Fan Yang and Zenan Zhou and Weipeng Chen},
      year={2025},
      eprint={2503.19470},
      archivePrefix={arXiv},
      primaryClass={cs.AI},
      url={https://arxiv.org/abs/2503.19470}, 
}

@misc{autorefine,
      title={Search and Refine During Think: Facilitating Knowledge Refinement for Improved Retrieval-Augmented Reasoning}, 
      author={Yaorui Shi and Sihang Li and Chang Wu and Zhiyuan Liu and Junfeng Fang and Hengxing Cai and An Zhang and Xiang Wang},
      year={2025},
      eprint={2505.11277},
      archivePrefix={arXiv},
      primaryClass={cs.CL},
      url={https://arxiv.org/abs/2505.11277}, 
}

@article{qwen3,
    title={Qwen3 Technical Report}, 
    author={An Yang and Anfeng Li and Baosong Yang and Beichen Zhang and Binyuan Hui and Bo Zheng and Bowen Yu and Chang Gao and Chengen Huang and Chenxu Lv and Chujie Zheng and Dayiheng Liu and Fan Zhou and Fei Huang and Feng Hu and Hao Ge and Haoran Wei and Huan Lin and Jialong Tang and Jian Yang and Jianhong Tu and Jianwei Zhang and Jianxin Yang and Jiaxi Yang and Jing Zhou and Jingren Zhou and Junyang Lin and Kai Dang and Keqin Bao and Kexin Yang and Le Yu and Lianghao Deng and Mei Li and Mingfeng Xue and Mingze Li and Pei Zhang and Peng Wang and Qin Zhu and Rui Men and Ruize Gao and Shixuan Liu and Shuang Luo and Tianhao Li and Tianyi Tang and Wenbiao Yin and Xingzhang Ren and Xinyu Wang and Xinyu Zhang and Xuancheng Ren and Yang Fan and Yang Su and Yichang Zhang and Yinger Zhang and Yu Wan and Yuqiong Liu and Zekun Wang and Zeyu Cui and Zhenru Zhang and Zhipeng Zhou and Zihan Qiu},
    journal = {arXiv preprint arXiv:2505.09388},
    year={2025}
}

@article{qwen2.5,
    title   = {Qwen2.5 Technical Report}, 
    author  = {An Yang and Baosong Yang and Beichen Zhang and Binyuan Hui and Bo Zheng and Bowen Yu and Chengyuan Li and Dayiheng Liu and Fei Huang and Haoran Wei and Huan Lin and Jian Yang and Jianhong Tu and Jianwei Zhang and Jianxin Yang and Jiaxi Yang and Jingren Zhou and Junyang Lin and Kai Dang and Keming Lu and Keqin Bao and Kexin Yang and Le Yu and Mei Li and Mingfeng Xue and Pei Zhang and Qin Zhu and Rui Men and Runji Lin and Tianhao Li and Tingyu Xia and Xingzhang Ren and Xuancheng Ren and Yang Fan and Yang Su and Yichang Zhang and Yu Wan and Yuqiong Liu and Zeyu Cui and Zhenru Zhang and Zihan Qiu},
    journal = {arXiv preprint arXiv:2412.15115},
    year    = {2024}
}

@misc{deepseekv3,
      title={DeepSeek-V3 Technical Report}, 
      author={DeepSeek-AI},
      year={2024},
      eprint={2412.19437},
      archivePrefix={arXiv},
      primaryClass={cs.CL},
      url={https://arxiv.org/abs/2412.19437}, 
}

@article{browsecomp-zh,
  title={BrowseComp-ZH: Benchmarking Web Browsing Ability of Large Language Models in Chinese},
  author={Zhou, Peilin and Leon, Bruce and Ying, Xiang and Zhang, Can and Shao, Yifan and Ye, Qichen and Chong, Dading and Jin, Zhiling and Xie, Chenxuan and Cao, Meng and others},
  journal={arXiv preprint arXiv:2504.19314},
  year={2025}
}

@inproceedings{webdancer,
title={WebDancer: Towards Autonomous Information Seeking Agency},
author={Jialong Wu and Baixuan Li and Runnan Fang and Wenbiao Yin and Liwen Zhang and Zhenglin Wang and Zhengwei Tao and Ding-Chu Zhang and Zekun Xi and Xiangru Tang and Yong Jiang and Pengjun Xie and Fei Huang and Jingren Zhou},
booktitle={The Thirty-ninth Annual Conference on Neural Information Processing Systems},
year={2025},
url={https://openreview.net/forum?id=quJdphBcdP}
}

@article{musique,
    title = "{M}u{S}i{Q}ue: Multihop Questions via Single-hop Question Composition",
    author = "Trivedi, Harsh  and
      Balasubramanian, Niranjan  and
      Khot, Tushar  and
      Sabharwal, Ashish",
    editor = "Roark, Brian  and
      Nenkova, Ani",
    journal = "Transactions of the Association for Computational Linguistics",
    volume = "10",
    year = "2022",
    address = "Cambridge, MA",
    publisher = "MIT Press",
    url = "https://aclanthology.org/2022.tacl-1.31/",
    doi = "10.1162/tacl_a_00475",
    pages = "539--554"
}

@inproceedings{bamboogle,
    title = "Measuring and Narrowing the Compositionality Gap in Language Models",
    author = "Press, Ofir  and
      Zhang, Muru  and
      Min, Sewon  and
      Schmidt, Ludwig  and
      Smith, Noah  and
      Lewis, Mike",
    editor = "Bouamor, Houda  and
      Pino, Juan  and
      Bali, Kalika",
    booktitle = "Findings of the Association for Computational Linguistics: EMNLP 2023",
    month = dec,
    year = "2023",
    address = "Singapore",
    publisher = "Association for Computational Linguistics",
    url = "https://aclanthology.org/2023.findings-emnlp.378/",
    doi = "10.18653/v1/2023.findings-emnlp.378",
    pages = "5687--5711"
}

@misc{llmasajudge,
      title={A Survey on LLM-as-a-Judge}, 
      author={Jiawei Gu and Xuhui Jiang and Zhichao Shi and Hexiang Tan and Xuehao Zhai and Chengjin Xu and Wei Li and Yinghan Shen and Shengjie Ma and Honghao Liu and Saizhuo Wang and Kun Zhang and Yuanzhuo Wang and Wen Gao and Lionel Ni and Jian Guo},
      year={2025},
      eprint={2411.15594},
      archivePrefix={arXiv},
      primaryClass={cs.CL},
      url={https://arxiv.org/abs/2411.15594}, 
}

@article{xVerify,
      title={xVerify: Efficient Answer Verifier for Reasoning Model Evaluations}, 
      author={Ding Chen and Qingchen Yu and Pengyuan Wang and Wentao Zhang and Bo Tang and Feiyu Xiong and Xinchi Li and Minchuan Yang and Zhiyu Li},
      journal={arXiv preprint arXiv:2504.10481},
      year={2025},
}

@inproceedings{dpr,
    title = "Dense Passage Retrieval for Open-Domain Question Answering",
    author = "Karpukhin, Vladimir  and
      Oguz, Barlas  and
      Min, Sewon  and
      Lewis, Patrick  and
      Wu, Ledell  and
      Edunov, Sergey  and
      Chen, Danqi  and
      Yih, Wen-tau",
    editor = "Webber, Bonnie  and
      Cohn, Trevor  and
      He, Yulan  and
      Liu, Yang",
    booktitle = "Proceedings of the 2020 Conference on Empirical Methods in Natural Language Processing (EMNLP)",
    month = nov,
    year = "2020",
    address = "Online",
    publisher = "Association for Computational Linguistics",
    url = "https://aclanthology.org/2020.emnlp-main.550/",
    doi = "10.18653/v1/2020.emnlp-main.550",
    pages = "6769--6781"
}

@misc{e5,
      title={Multilingual E5 Text Embeddings: A Technical Report}, 
      author={Liang Wang and Nan Yang and Xiaolong Huang and Linjun Yang and Rangan Majumder and Furu Wei},
      year={2024},
      eprint={2402.05672},
      archivePrefix={arXiv},
      primaryClass={cs.CL},
      url={https://arxiv.org/abs/2402.05672}, 
}

@misc{asearcher,
      title={Beyond Ten Turns: Unlocking Long-Horizon Agentic Search with Large-Scale Asynchronous RL}, 
      author={Jiaxuan Gao and Wei Fu and Minyang Xie and Shusheng Xu and Chuyi He and Zhiyu Mei and Banghua Zhu and Yi Wu},
      year={2025},
      eprint={2508.07976},
      archivePrefix={arXiv},
      primaryClass={cs.CL},
      url={https://arxiv.org/abs/2508.07976}, 
}

@misc{mirorl,
    title={MiroRL: An MCP-first Reinforcement Learning Framework for Deep Research Agent},
    author={ MiroMind Foundation Model Team and MiroMind AI Infra Team},
    howpublished = {\url{https://github.com/MiroMindAI/MiroRL}},
    year={2025}
}
\appendix

\end{document}